# Demonstration of low power and highly uniform 6-bit operation in SiO$_2$-based memristors embedded with Pt nanoparticles

G. Kleitsiotis, P. Bousoulas, S. D. Mantas, C. Tsioustas, I. A. Fyrigos, G. Sirakoulis, D. Tsoukalas

*Abstract*— **In this work, an optimized method was implemented for attaining stable multibit operation with low energy consumption in a two-terminal memory element made from the following layers: Ag/Pt nanoparticles (NPs)/SiO$_2$/TiN in a 1-Transistor-1-Memristor configuration. Compared to the reference sample where no NPs were embedded, an enlarged memory window of ~$10^5$ was recorded in conjunction with reduced variability for both switching states ($\sigma/\mu$ <0.4). A comprehensive numerical model was also applied to shed light on this enhanced performance, which was attributed to the spatial confinement effect induced by the presence of the Pt NPs and its impact on the properties of the percolating conducting filaments (CFs). Although 5-bit precision was demonstrated with the application of the incremental-step-pulse-programming (ISPP) algorithm, the reset process was unreliable and the output current increased abnormally when exceeded the value of 150 μA. As a result, the multibit operation was limited. To address this issue, a modified scheme was developed to accurately control the distance between the various resistance levels and achieve highly reliable 6-bit precision. Our work provides valuable insights for the development of energy-efficient memories for applications where a high density of conductance levels is required.**

## I. INTRODUCTION

THE constant increase in the amount of generated data in the Internet of Things era requires the development of new storage methods and reconfigurable computational systems. The latter should have the ability to perform high-speed processing at the edge in an energy-efficient manner [1]. To effectively address these challenges, cutting-edge innovations towards non-conventional computer architectures are required. The various emergent compute-in-memory (CIM) technologies surpass conventional systems in solving complex tasks due to their inherent parallel processing properties. The current state-of-the-art conventional systems rely on V-NAND class flash storage drives for achieving sustainable non-volatile data storage. The latter display excellent bit density due to their 3-dimensional stack design, as well as the novel multistate cell architectures. As a result, a V-NAND configuration could read/write data at the granularity of the page. However, the erase process requires copying "to be preserved" data to other pages. Moreover, the erase of entire blocks of memory induces a complex and computationally intense procedure that often leads to data corruption issues due to repeated write/erase sequences. Due to significant hardware restrictions, it is also almost impossible to process the stored amount of data locally, relying on remote "cloud computing" methods, further increasing the operating cost, in terms of energy consumption and latency due to their constant data transfer requirement [2].

On the contrary, within a crossbar array architecture (CBA) composed of memory elements with tunable resistance levels, similar to the NOR flash architecture, random access operation of each individual cell is feasible [3], [4]. Due to their analogue operation nature, not only do they offer a low-consumption storage alternative [5], but also allow instantaneous CIM execution of the multiply and accumulate operations, in accordance with Ohm and Kirchhoff's laws, respectively [6]. Normally, a 1-Transistor-1-Memristor (1T1M) architecture is selected to deal with the significant disturbances of the sneak path current in a CBA configuration [7].

Among the various non-volatile memory concepts, resistive switching memories (RRAMs) arise as a promising candidate due to their ultra fast switching dynamics, simple structure, and low power consumption [8]. Their operation relies upon the transition from a high- to low-resistance state (HRS & LRS), vice and versa, under the application of a voltage difference between two electrodes. Considering that the 1T1M implementation of RRAMs has a similar cell area to the NAND flash memories, an increased storage density is considered of great importance for the realistic replacement of the existing memory technologies [9]. However, exceptional fine-tuning is required to achieve the required resistance states, which is challenging due to the stochastic nature of the resistive switching effect [10]. As a result, a direct control of the defect density within RRAMs is required to tune the properties of the formed conducting filament (CF), which is not always feasible due to the utilization of complicated fabrication processes or energy-inefficient electroforming procedures [11].

It is thus apparent that optimized programming schemes are required to leverage the unique properties of RRAMs. For this reason, various programming algorithms based on the application of the industrial compatible incremental-step-pulse-programming (ISPP) method have been proposed in the literature [12] – [15]. However, all the reported schemes are applicable to devices with small memory windows, which could create current overshoots when being applied to other types of devices. A modified method is thus required to permit the accurate programming of the device on the desired resistance state and leverage the whole resistance range. To this end, in this work, an adaptive state control algorithm (ASCA) was developed for attaining reliable multibit operation. The underlying idea of our approach is to exploit the gradual switching transitions of the fabricated memory elements, which render them more tolerant to voltage variations, exhibiting large memory windows. To achieve this goal, the device operation was divided into discrete conductivity bands, where appropriate pulsing schemes were applied. As a result, the density of the multiple resistance states was substantially increased (6-bit) in striking contrast with the conventional ISPP scheme (5-bit), where uncontrollable patterns at high current values (>200 μA) were recorded. In parallel, a Pt nanoparticles (NPs)-based memory was introduced to increase the available memory window and reduce the variability of the various resistance levels. The latter was also obtained with the utilization of a 1T1M configuration for attaining better control over the various complicate limit (I$_{CC}$) currents. Finally, the prevailing physical mechanisms of the switching effect were analyzed with the application of a comprehensive numerical model.

## II.  MATERIALS AND METHODS

Two samples were fabricated to examine their properties as conductive bridge memories (CBRAM). More specifically, the reference sample (Sample R) comprises an electrochemically inert TiN bottom electrode (BE), a SiO$_x$ dielectric (active) layer, and an electrochemically active top electrode (TE) of Ag. The NP-enhanced sample (Sample NPs) includes one layer of Pt NPs beneath the TE. The 1T1M configuration was completed by connecting the drain terminal of a BS170 n-MOSFET to the BE of the CBRAM. All layers of the CBRAM cell were deposited at a final thickness of 40 nm by RF magnetron sputtering. The chamber was dropped at a pressure of $2 \times 10^{-6}$ mbar, before the Ar plasma was ignited at a stable pressure of $4 \times 10^{-3}$ mbar and a total power of 150 W. High-purity Ag and TiN targets for TE and BE were used, respectively. Square TEs with 100 μm dimensions were fabricated. Reactive RF magnetron sputtering was used similarly to deposit the 40 nm thick oxide layer, using a 99.99% pure ceramic target, inside a 20:1.5 (Ar:O$_2$) environment at $4 \times 10^{-5}$ mbar. The Pt NPs layer was formed by performing direct current (DC) magnetron sputtering, using a nanoparticle generator (NanoGen). By sputtering Pt inside a small chamber through an aperture of approximately 5 nm, at a chamber pressure of $2.2 \times 10^{-3}$ mbar and a DC current of 0.1 A, and setting the condensation length at 20 cm, NP aggregates of 3 nm average diameter were deposited at a final density of approximately $2 \times 10^{12}$ NCs/cm². The latter was extracted by performing transmission electron microscopy (TEM) measurements.

A 3-terminal set-up was used in a SUSS MicroTech probe station to evaluate the electrical performance of the 1T1M configuration. 2 terminals (channels) were responsible for biasing the TE and adjusting the MOSFET gate. The third terminal acted as a common channel connected to the transistor source. The DC characteristics were extracted using a Keithley 4200 semiconductor characterization system (SCS). Pulse-bias measurements were conducted through the Keithley 4201 pulse measurement unit along with the Keithley 4225 remote sensing amplifiers.

## III.  ELETRICAL PERFROMANCE EVALUATION

### A.  DC analysis and numerical modelling

Fig. 1 depicts the cross-section of the fabricated samples, as well as the recorded hysteresis patterns. No forming process was applied before to device operation and a stable bipolar switching pattern was obtained for both cases. The onset of the SET transition for Sample R takes place continuously from 0.3 V, whereas for Sample NPs, the same process occurs at about 0.5 V. This behavior originates from the spatial confinement effect induced by the presence of Pt NPs [16]. As a result, the potential available paths for the creation of the conducting filament (CF) are reduced. On the other hand, the effect is desirable for enhancing the intrinsic variability of the memristive elements. The RESET process takes place for both samples at about -0.3 V with an abrupt pattern, indicating the manifestation of a thermal-based mechanism.

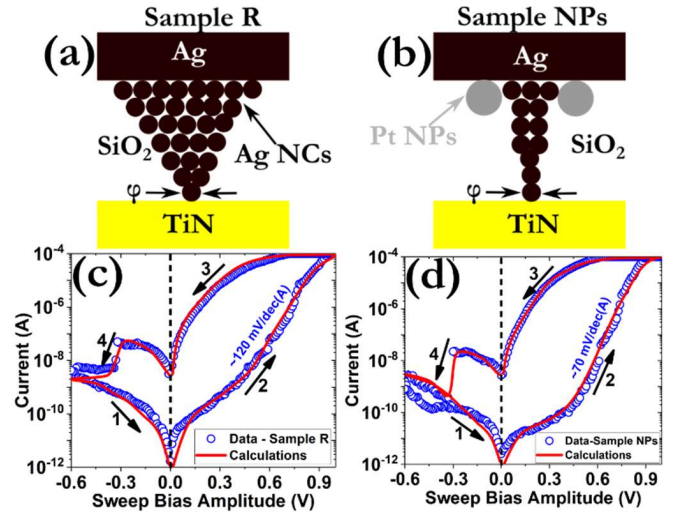

Fig. 1.  Schematic representation of the cross-section of (a) Sample R and (b) Sample NPs. Measured and simulated hysteresis curves for (c) Sample R and (d) Sample NPs with a sweep rate of 10 mV/s. A constant I$_{CC}$ of 100 μA was also enforced for all samples.

To better understand the origins of the above-mentioned experimental results, as well as the driving mechanisms responsible for CF creation/rupture processes, a previously reported self-consistent numerical model was applied [17], [18]. The existence of CF is attributed to the independence of the LRS from the total device area (not shown here). The CF was also assumed to consist of Ag ions exhibiting metallic properties, as was extracted by electrical temperature measurements at high temperatures (not shown here). Under the application of external signals, metal nanoclusters (NCs) composed of Ag are progressively formed as a result of the accelerated material precipitation because of the increased solid-solubility of Ag [19]. In addition, truncated-coned CFs were assumed to deal with the self-rectification pattern of the fabricated devices. This conjecture is in direct line with concrete experimental pieces of evidence concerning SiO₂-based memristors [20], [21]. The whole memristive behavior can be examined by calculating the CF's effective diameter (φ) by solving the following differential equations:

$$\frac{d\varphi}{dt} = \frac{d\varphi}{dt}\bigg|_{drift} + \frac{d\varphi}{dt}\bigg|_{diffusion} + \frac{d\varphi}{dt}\bigg|_{thermo-diffusion} =$$

$$= Ae^{\frac{E_{drift}-aq\psi}{k_BT}} + B\varphi^{-1}e^{\frac{E_{diff}}{k_BT}} - C\varphi^{-1}S\left(\frac{\partial T}{\partial r}+\frac{\partial T}{\partial z}\right) \quad (1)$$

$$\nabla \cdot \sigma \nabla \psi = 0 \quad (2)$$

$$\rho_m C_p \frac{\partial T}{\partial t} = \nabla k_{th} \cdot \nabla T + \sigma \nabla |\psi|^2 \quad (3)$$

where $k_B$ is the Boltzmann constant, T is the absolute temperature, α is the barrier lowering factor, $E_{drift}$ and $E_{diff}$ are the energy barriers for ion hopping and diffusion respectively, ψ is the electrical potential, σ is the electrical conductivity, $E_S$ is the activation energy of thermophoresis, S is the Soret coefficient, $k_{th}$ is the thermal conductivity, $\rho_m$ is the mass density, $C_p$ is specific heat, and A, B, and C are constants. The specific values of all parameters can be found elsewhere [5], [17], [18]. The set of equations was solved self-consistently and simultaneously using a numerical solver (COMSOL).

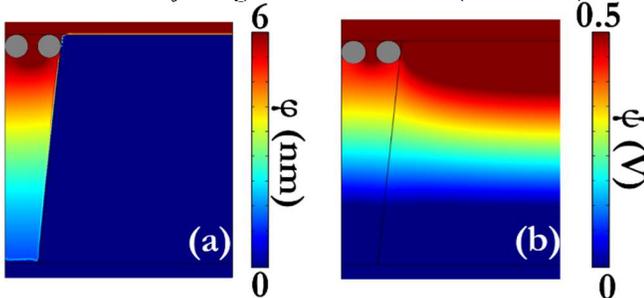

Fig. 2. 3D calculated maps of (a) the CF's effective diameter and (b) localized temperature distribution during the SET transition.

Truncated-cone CFs were drawn to deal with the self-rectification effect, which could not be explained taking into account the Schottky barrier height difference between the TE and the oxide, as well as between the oxide and the BE. For both samples, a CF with the same value at the interface with the BE was used (6 nm diameter). This selection was made since both samples exhibit the same current values during the SET process (Fig. 1(c), (d)). However, to account for the steeper switching slope of Sample NPs, smaller values for the ionic migration barriers through drift and diffusion mechanisms were employed ($E_{drift}$ = 1 eV and $E_{diff}$ = 1.1 eV for Sample NPs and $E_{drift}$ = 1.3 eV and $E_{diff}$ = 1.25 eV for Sample R). These values are also compatible with the underlying physical mechanism induced by the presence of the Pt NPs layer. More specifically, the spatial confinement reduces the available degrees of freedom for the movement of silver ions. As a result, an increased density of silver ions is anticipated in the region between the NPs (Fig. 2(a)), as well as an increased voltage drop in this region (Fig. 2(b)).

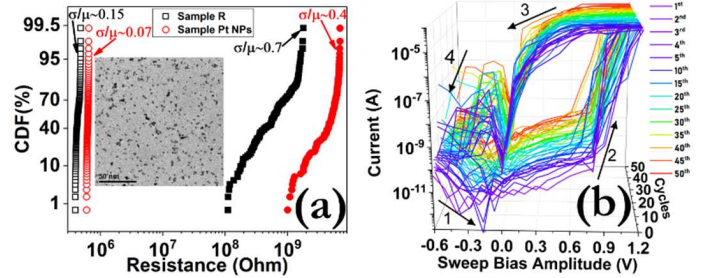

Fig. 3. (a) Cumulative distribution functions (CDFs) for both samples. The open symbols correspond to the LRS and the filled ones to the HRS. The data have been collected by measuring 150 different devices on each sample. The inset depicts a TEM of the Pt NPs layer, where the scale bar corresponds to 50 nm. (b) I-V consecutive cycles after the application of 50 DC endurance cycles for Sample NPS.

The statistical dispersion of the device-to-device and cycle-to-cycle characteristics were also investigated. As can be observed from Fig. 3(a), the coefficient of variance (σ/μ) of Sample NPs was improved with respect to the reference sample. This enhancement stems from the controllable onset of the SET transition, which is anticipated due to the presence of Pt NPs beneath the TE. Although the variability of the HRS was reduced for Sample NPs, the respective value of the σ/μ coefficient is still relatively high. This effect could be explained by considering that the RESET process takes place at the interface of the tip of the CF and the BE. As a result, it cannot be directly tuned. This pattern is also discernible during the DC endurance measurements, where the RESET transition displays a bigger variation. We have to underline that the diverse sizes of the Ag NCs also affect the dispersion of the switching characteristics since they will possess diverse diffusivity barriers and activation energies. Consequently, the percolating CF will each time be generated and annihilated at various voltages and a different feedback will be provided by the competition of the drift, diffusion, and thermos-diffusion fluxes. In any case, the performance of Sample NPs is considered satisfactory since the initial memory window is preserved.

## B. AC analysis

The switching kinetics was also explored by varying the amplitude of the SET pulse, as can be seen in Fig. 4. Particularly, square pulses with a fixed width of 1 μs and a rise time of 0.1 μs were applied. A direct relationship between the amplitude and the SET switching time was extracted, yielding a relatively steep slope of about 110 ns/V. The high non-linearity of the SET kinetic process can be ascribed to the local temperature and the field-accelerated migration of silver ions, which underscores the critical effect of these two driving forces. Interestingly, for Samples R, smaller slopes were calculated (not shown here), clearly illustrating the beneficial role of the one layer of Pt NCs on the robust and controllable evolution of the memristive effect.

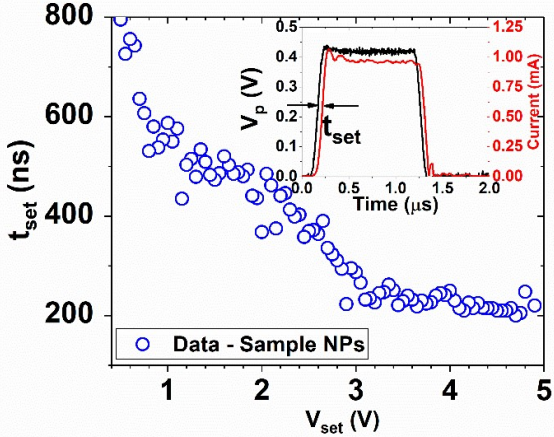

Fig. 4. Dependence of the SET time ($t_{SET}$) as a function of the pulse amplitude, indicating the high nonlinearity nature of the SET kinetics process. The inset depicts a pulsed I-V characteristic.

## IV. APPLICATION OF THE ISPP SCHEME

To leverage the gradual switching transitions of the Sample NPs and the large memory window, the implementation of the conventional ISPP protocol that is widely used in flash memories was initially examined [22]. A selection transistor was used to enforce an upper limit to the current flowing through the memristor, which was connected in series with the transistor's drain (Fig. 5(a)). The current compliance limit was attained under the application of appropriate bias to the gate terminal of the transistor ($V_{GATE}$). To achieve stable retention performance, a minimum of $I_{CC}$ of 10 µA was selected, while the maximum available current was 250 µA. At the same time, square pulses with 0.5 ms width and increasing amplitudes from 0 to 3 V were applied to the memristor's TE. The read-out process was performed by applying 200 mV square pulses with the same width after each programming step and keeping the $I_{CC}$ at its upper limit. Although with the application of this scheme, 5-bit multilevel capabilities were achieved [23], [24], it was not feasible to keep the distance between the various discrete levels constant. The latter effect obviously leads to a reduced number of potentially available states. However, the main disadvantage of this approach lies in the failure of the device after conducting consecutive programming. More specifically, in most of the cases (at least 100 different devices were tested), we were not able to fully erase the device and reduce the operating current values.

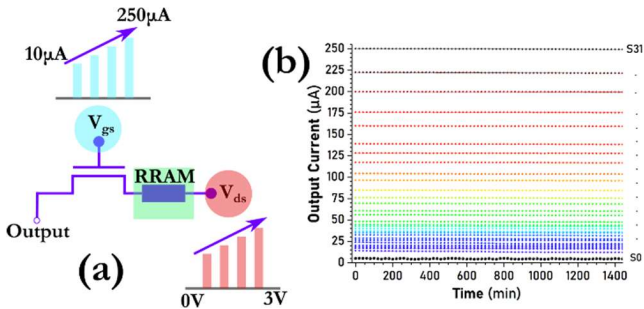

Fig. 5. (a) Schematic illustration of the ISPP scheme in a 1T1M configuration. (b) Retention performance for 32 discrete conductivity levels (5-bit performance).

As a result, this method was proven inadequate for our prototypes in order to demonstrate robust and reliable multilevel capabilities. This phenomenon is directly related to the uncontrollable nature of the CF, which raises reliability issues during device operation under strict programming conditions. It is thus apparent that the development of new programming algorithms is required.

## V. APPLICATION OF THE ASCA SCHEME

To effectively address the above-mentioned issues, a more precise process to tune the various conductance levels was devised. To accomplish this, the ASCA scheme was employed (Fig. 6). Particularly, during the application of ASCA, the desired target state of the device is divided into three separate conductivity bands. In each band, separate programming schemes are applied to accurately program the device (Fig. 7(b)). The underlying concept here is that after each SET and RESET pulse, read-out pulses are always enforced to confirm whether the output current lies within a target current range with a specified 5% tolerance. If the output current is higher, RESET pulses are applied and the device is re-programmed with a new configuration of SET pulses. In the case of a lower output current, the programming procedure is ended. After the successful programming of each state, an erase scheme is also always enforced.

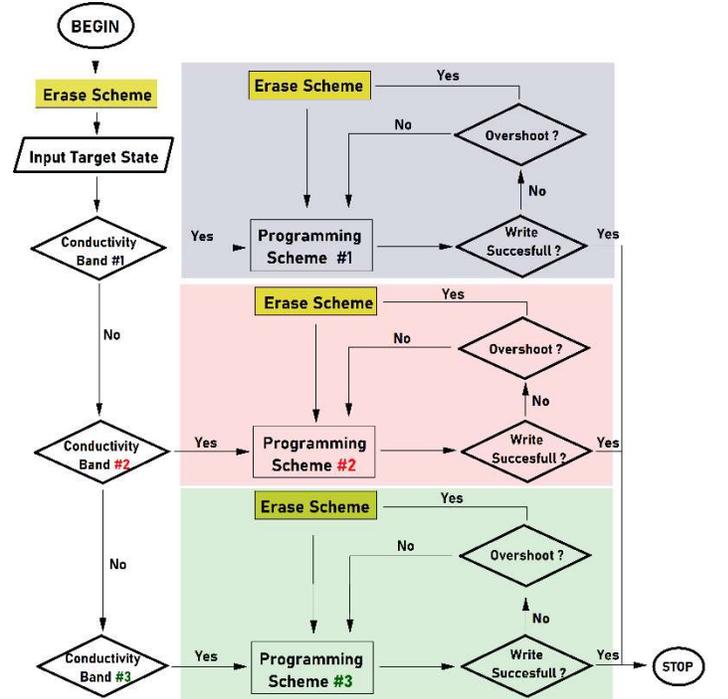

Fig. 6. Flowchart presentation of the ASCA scheme.

Fig. 7 presents the various applied schemes that were used for the programming and erase processes of our prototypes. As the erase scheme is concerned (Fig. 7(a)), a constant $V_{GATE}$ of 1.85 V was applied to the transistor to keep it constantly open and initially, a series of 60 negative pulses with an amplitude of -0.5 V and width of 0.5 ms were applied. This number of pulses was selected to attain a full RESET of the device and reduce the operating currents below 10 µA. If this not achieved, a second erase protocol was applied, where the amplitudes were incremented until the termination condition was met. After each erase pulse, in the second case, a read-out pulse is also enforced (200 mV amplitude). For the programming scheme, a more sophisticated approach was used (Fig. 7(b)). When the operating current of the devices after the initial read-out process is detected in the first current zone ($10 - 50$ µA), the device can be written in a specific state simply by applying a series of increasing voltage pulses from 1.5 to 3 V and 1 ms width, with the gate biased at a constant voltage value, predefined for each discrete level, thus defining an upper limit to current flow. To verify successful programming, a read-out pulse is applied (200 mV), with the gate biased at 3.0 V to avoid possible false readings caused by the saturating n-MOSFET. If the read-out current is below the targeted state, the algorithm proceeds first with the application of the next pulse in the series ($V_{DRAIN}$) and then the $V_{GATE}$ is further increased and the whole sequence is repeated. With the detection of an overshoot, where the conductivity is above the predefined 5% tolerance, the device is erased and the process will begin again, with the amplitude parameters of the pulse train (voltage and step increment) adjusted accordingly. Once the first current zone has been populated, the programming process is continued in the second zone ($50 - 120$ µA), where a different scheme is now used. More specifically, read-out feedback is received every three applied pulses to the drain, and a similar Scheme 1 parameter adjustment is followed in the event of an overshoot. Additionally, the process continues by raising the amplitude of the $V_{GATE}$ to meet the target conductivity level. Finally, for read-out currents exceeding the threshold of 120 µA, Scheme 3 is applied where the drain is pulsed similarly to the case of Scheme 2, however, the gate bias is incremented after each stimulus is applied to the drain. The process ends when the target conductivity state is reached. Then, the device is erased, the parameters are adjusted, and the process is restarted with the detection of an overshoot.

three different pulse sequences are enforced, namely Schemes 1, 2 and 3.

From a physical point of view, the diverse sequence of pulses between Scheme 1 and Schemes 2 & 3 reflects the different dynamics during the CF formation process [25], [26]. Particularly, at low operating currents, the CF is mainly vertically grown and affected by the rapid migration of Ag ions through the drift effect. As a result, an incremental-based programming protocol can be applied without affecting the density of the induced conductivity states. Once the CF has been formed, the diffusion effect dominates, which leads to lateral enlargement of the CF's diameter. Now, the device is more susceptible to the applied programming procedure. For this reason, a set of 3 programming pulses was experimentally proven beneficial to stabilize the induced conductance state and leverage the whole available spectrum.

The devices were also tested, in terms of memory retention, using artificial ageing with a tolerance of 5% variation for each initial conductivity level. After being set to their respective state, they were tested at 150 ºC for 24 hours, inside a vacuum chamber and with consecutive read-outs being conducted during the process. With the application of the ASCA scheme, Sample R yielded 4-bit performance and Sample NPs 6-bit (Fig. 8). The cycle-to-cycle variation for the latter sample was also examined after being programmed a total of 50 times to each of the 63 LRS, demonstrating outstanding variability performance (Fig. 8). The proposed cell architecture is also promising for implementing quantum computer accelerators with increased mapping accuracy of the quantum gates [27], [28].

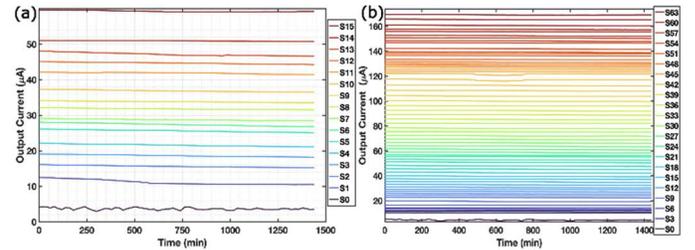

Fig. 8. Retention performance for (a) Sample R displaying 16 discrete conductivity levels (4-bit performance) and (b) Sample NPs displaying 64 discrete conductivity levels (6-bit performance).

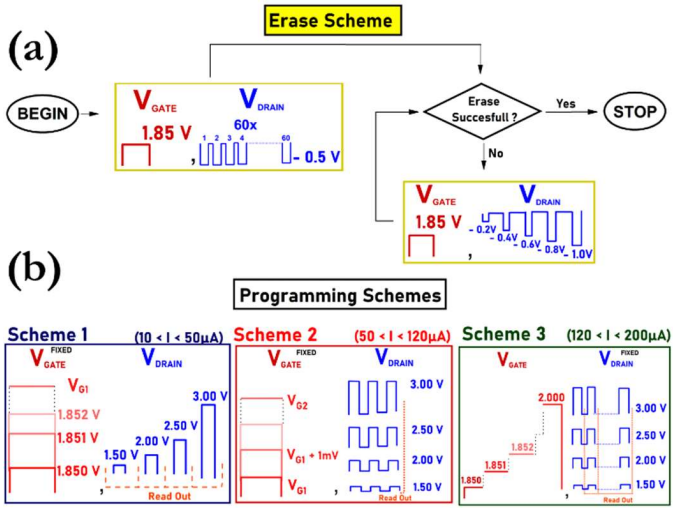

Fig. 7. Schematic representation of the applied schemes. (a) During the application of the erase scheme, a middle reset procedure is initially applied, followed by the application of an incrementing negative bias until the device reaches its HRS. (b) In the programming scheme,

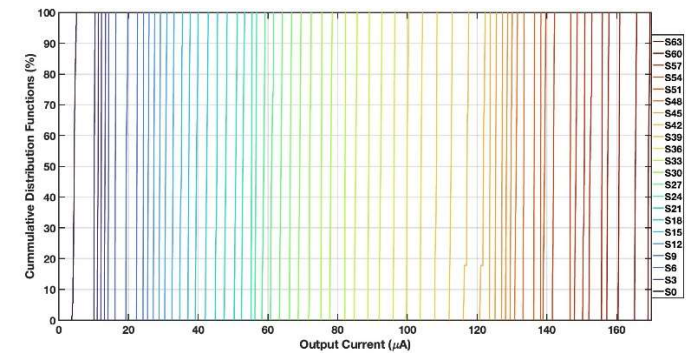

Fig. 9. Distribution of the variability of all conductivity states during cycle-to-cycle operation (Sample NPs).

## VI. CONCLUSION

In this work, an optimized scheme was proposed to leverage the gradual switching patterns of the proposed device combined with the high memory window and reduced variability. The latter was attained by incorporating a layer of Pt NCs beneath the TE of the CBRAM device. As a result of this optimized device configuration, 6-bit performance was demonstrated with low power consumption by using the proposed ASCA scheme, while with the application of the standard ISSP scheme, 5-bit performance was extracted. A comprehensive numerical model was also applied to elucidate the origins of the enhanced switching effect. Our work paves the way for the development of robust and low-power memory elements with enhanced multibit capabilities.

## REFERENCES

[1] C. Wang, Z. Si, X. Jiang, A. Malik, Y. Pan, S. Stathopoulos, A. Serb, S. Wang, T. Prodromakis, C. Papavassiliou, "Multi-State Memristors and Their Applications: An Overview", vol. 12, no. 4, p.p. 723 – 734, *IEEE Journal on Emerging and Selected Topics in Circuits and Systems*, Dec. 2022, doi: 10.1109/JETCAS.2022.3223295.

[2] H. Xue, F. Dai, G. Liu, P. Cao, and B. Huang, "Edge Computing: A Systematic Mapping Study," in 2021 IEEE Intl Conf. on Dependable, Autonomic and Secure Computing, Intl Conf on Pervasive Intelligence and Computing, Intl Conf on Cloud and Big Data Computing, Intl Conf. on Cyber Science and Technology Congress (DASC/PiCom/CBDCom/CyberSciTech), Oct. 2021, pp. 507-514. doi: 10.1109/DASC-PICom-CBDCom-CyberSciTech52372.2021.00090.

[3] S. S. Kim et al., "Review of Semiconductor Flash Memory Devices for Material and Process Issues," *Advanced Materials*, vol. 35, no. 43, p.p. 2200659, Oct. 2023. doi: 10.1002/adma.202200659.

[4] K. J. Yoon, Y. Kim, and C. S. Hwang, "What Will Come After V-NAND—Vertical Resistive Switching Memory?," *Advanced Electronic Materials*, vol. 5, no. 9, p. 1800914, Sept. 2019. doi: 10.1002/aelm.201800914.

[5] P. Bousoulas, C. Tsioustas, J. Hadfield, V. Aslanidis, S. Limberopoulos, and D. Tsoukalas, "Low Power Stochastic Neurons From SiO₂-Based Bilayer Conductive Bridge Memristors for Probabilistic Spiking Neural Network Applications—Part II: Modeling," *IEEE Transactions on Electron Devices*, vol. 69, no. 5, pp. 2368-2376, May 2022. doi: 10.1109/TED.2022.3160140.

[6] S. Jung, H. Lee, S. Myung, et al., "A crossbar array of magnetoresistive memory devices for in-memory computing," *Nature*, vol. 601, no. 7892, pp. 211-216, Jan. 2022. doi: 10.1038/s41586-021-04196-6.

[7] V. Sakode, F. Lombardi, and J. Han, "Cell design and comparative evaluation of a novel 1T memristor-based memory," in *2012 IEEE/ACM International Symposium on Nanoscale Architectures (NANOARCH)*, Jul. 2012, pp. 152–159. doi: 10.1145/2765491.2765519.

[8] U. Böttger, M. von Witzleben, V. Havel, K. Fleck, V. Rana, R. Waser, S. Menzel, "Picosecond multilevel resistive switching in tantalum oxide thin films," *Sc. Reports*, vol. 10, no. 16931, Oct. 2020, doi: s41598-020-73254-2.

[9] P. Bousoulas, S. Stathopoulos, D. Tsialoukis, D. Tsoukalas, "Low-Power and Highly Uniform 3-b Multilevel Switching in Forming Free TiO₂-ₓ-Based RRAM With Embedded Pt Nanocrystals, " vol. 37, no. 7, p.p. 874 – 877, *IEEE Electron Device Lett.*, July 2016, doi: 10.1109/LED.2016.2575065.

[10] A. Prakash, J. Park, J. Song, J. Woo, E.-J. Cha, H. Hwang, "Demonstration of Low Power 3-bit Multilevel Cell Characteristics in a TaOₓ-Based RRAM by Stack Engineering," *IEEE Electron Device Letters*, vol. 36, no. 1, pp. 32-34, Jan. 2015. doi: 10.1109/LED.2014.2375200.

[11] W. Banerjee and H. Hwang, "Quantized Conduction Device with 6-Bit Storage Based on Electrically Controllable Break Junctions," *Advanced Electronic Materials*, vol. 5, no. 12, Dec. 2019. doi: 10.1002/aelm.201900744.

[12] E. Pérez, C. Zambelli, M. K. Mahadevaiah, P. Olivo, C. Wegner, "Toward Reliable Multi-Level Operation in RRAM Arrays: Improving Post-Algorithm Stability and Assessing Endurance/Data Retention," *IEEE Journal of the Electron Devices Society*, vol. 7, p.p. 740-747, July 2019. doi: 10.1109/JEDS.2019.2931769.

[13] S. Stathopoulos, A. Khiat, M. Trapatseli, S. Cortese, A. Serb, I. Valov, T. Prodromakis, "Multibit memory operation of metal-oxide bi-layer memristors," *Scientific Reports*, vol. 7, no. 1, Dec. 2017. doi: 10.1038/s41598-017-17785-1.

[14] G. H. Kim, H. Ju, M. K. Yang, D. K. Lee, J. W. Choi, J. H. Jang, S. G. Lee, I. S. Cha, B. K. Park, J. H. Han, T.-M. Chung, K. M. Kim, C. S. Hwang, Y. K. Lee, "Four-Bits-Per-Cell Operation in an HfO₂-Based Resistive Switching Device," *Small*, vol. 13, no. 40, p.p. 1701781, Oct. 2017. doi: 10.1002/smll.201701781

[15] J. J. Ryu, B. K. Park, T.-M. Chung, Y. K. Lee, G. H. Kim, "Optimized Method for Low-Energy and Highly Reliable Multibit Operation in a HfO₂-Based Resistive Switching Device," *Ad. Electron. Mater.*, vol. 4, no. 12, p.p. 1800261, Dec. 2018 doi: 10.1002/smll.201701781.

[16] D. Sakellaropoulos, P. Bousoulas, C. Papakonstantinopoulos, S. Kitsios, D. Tsoukalas, "Spatial Confinement Effects of Embedded Nanocrystals on Multibit and Synaptic Properties of Forming Free SiO₂-Based Conductive Bridge Random Access Memory, vol. 41, no. 7, p.p. 1013 – 1016, *IEEE Electron Device Lett.*, May 2020, doi: 10.1109/LED.2020.2997565.

[17] P. Bousoulas, D. Sakellaropoulos, C. Papakonstantinopoulos, S. Kitsios, C. Arvanitis, E. Bagakis, D. Tsoukalas, "Investigating the origins of ultra-short relaxation times of silver filaments in forming-free SiO₂-based conductive bridge memristors", vol. 31, no. 45, *Nanotechnology*, Aug. 2020, Art no. 454002, doi: 10.1088/1361-6528/aba3a1.

[18] P. Bousoulas, D. Sakellaropoulos, and D. Tsoukalas, "Tuning the analog synaptic properties of forming free SiO₂ memristors by material engineering," *Applied Physics Letters*, vol. 118, no. 14, p.p. 143502, Apr. 2021. doi: 10.1063/5.0044647.

[19] M.-L. Avramescu, P. E. Rasmussen, M. Chénier, and H. D. Gardner, "Influence of pH, particle size and crystal form on dissolution behavior of engineered nanomaterials," *Environ. Sci. Pollut. Res.*, vol. 24, no. 2, pp. 1553–1564, Jan. 2017. doi: 10.1007/s11356-016-7932-2.

[20] Y. Yang, P. Gao, S. Gaba, T. Chang, X. Pan, W. Lu, "Observation of conducting filament growth in nanoscale resistive memories", vol. 3, *Nat. Commun.*, March 2012, Art. no. 732 doi: 10.1038/ncomms1737.

[21] H. Sun, Q. Liu, C. Li, S. Long, H. Lv, C. Bi, Z. Huo, L. Li, M. Liu, "Direct Observation of Conversion Between Threshold Switching and Memory Switching Induced by Conductive Filament Morphology", vol. 24, no. 36, p.p. 5679 – 5686, *Adv. Mater.*, Sept. 2014, doi: 10.1002/adfm.201401304.

[22] K.-T. Park, M. Kang, S. Hwang, Y. Song, J. Lee, H. Joo, H.-S. Oh, J.-J. Kim, Y.-T. Lee, C. Kim, W. Lee, "Dynamic Vpass ISPP scheme and optimized erase Vth control for high program inhibition in MLC NAND flash memories," 2009 Symposium on VLSI Circuits, Kyoto, Japan, pp. 24-25, June 2009, doi: https://ieeexplore.ieee.org/stamp/stamp.jsp?tp=&arnumber=5205328&isnumber=5205281.

[23] F. G. Aga, J. Woo, J. Song, J. Park, S. Lim, C. Sung, H. Hwang, "Controllable quantized conductance for multilevel data storage applications using conductive bridge random access memory", vol. 28, p.p. 115707, *Nanotechnology*, Feb. 2017, doi: 10.1088/1361-6528/aa5baf.

[24] V. K. Sahu, A. K Das, R. S. Ajimsha, P. Misra, "Low power high speed 3-bit multilevel resistive switching in TiO₂ thin film using oxidisable electrode", vol. 53, p.p. 225303, *J. Phys. D: Appl. Phys.*, March 2020, doi: 10.1088/1361-6463/ab7acb.

[25] S. Yu and H.-S. P. Wong, "Compact modeling of conducting bridge random-access Memory (CBRAM)," *IEEE Transactions on Electron Devices*, vol. 58, no. 5, pp. 1352-1360, May 2011. doi: 10.1109/TED.2011.2116120.

[26] C. Tsioustas, P. Bousoulas, J. Hadfield, T. P. Chatzinikolaou, I.-A. Fyrigos, V. Ntinas, M.-A. Tsompanas, G. Ch. Sirakoulis, and D. Tsoukalas, "Simulation of low power self-selective memristive neural networks for in situ digital and analogue artificial neural network applications," *IEEE Transactions on Nanotechnology*, vol. 21, p.p. 505-513, May 2011. doi: 0.1109/TNANO.2022.3205698.

[27] I.-A. Fyrigos, V. Ntinas, N. Vasileiadis, Georgios Ch. Sirakoulis, P. Dimitrakis, Y. Zhang, I. G. Karaflyllidis, "Memristor Crossbar Arrays Performing Quantum Algorithms," IEEE Transactions on Circuits and Systems I:Regular Papers, vol. 69, no. 2, p.p. 552-563, Feb. 2022. doi: 10.1109/TCSI.2021.3123575.

[28] I.-A. Fyrigos, T. P. Chatzinikolaou, V. Ntinas, N. Vasileiadis, P. Dimitrakis, I. Karafyllidis, and G. Ch. Sirakoulis, "Memristor Crossbar Design Framework for Quantum Computing," in 2021 IEEE